\documentclass[10pt,conference]{IEEEtran}
\usepackage{algorithm,amsbsy,amsmath,amssymb,epsfig,bbm,mathrsfs,multirow,amsthm}
\usepackage{color}
\usepackage{algpseudocode,tabularx}
\usepackage{graphicx,caption,subcaption,array,multirow,bm}
\usepackage{makecell}
\usepackage{cite}
\usepackage{threeparttable}
\usepackage[printonlyused,withpage]{acronym}
\usepackage{adjustbox}
\usepackage{array}

\DeclareMathAlphabet{\mathpzc}{OT1}{pzc}{m}{it}

\makeatletter
\newcommand{\multiline}[1]{%
	\begin{tabularx}{\dimexpr\linewidth-\ALG@thistlm}[t]{@{}X@{}}
		#1
	\end{tabularx}
}
\makeatother

\newlength\mylength
\begin{document}

 \newacro{LTI}{Linear time-invariant}
 \newacro{3GPP}{Third Generation Partnership Project}
 \newacro{2G}{second-generation}
 \newacro{GSM}{global system for mobile communications}
 \newacro{EDGE}{enhanced data rates for global evolution}
 \newacro{IS-95}{interim standard 95}
 \newacro{3G}{third-generation}
 \newacro{4G}{fourth-generation}
 \newacro{5G}{fifth-generation}
 \newacro{mmWave}{millimeter wave}
 \newacro{6G}{sixth-generation}
 \newacro{LTE}{long-term evolution}
 \newacro{PSS}{primary synchronization signal}
 \newacro{SSS}{secondary synchronization signal}
 \newacro{TDMA}{time-division-multiple-access}
 \newacro{FDMA}{frequency division multiple access}
 \newacro{SC-FDMA}{single carrier frequency division multiple access}
 \newacro{CDMA}{code division multiple access}
 \newacro{PN-codes}{pseudo random codes}
 \newacro{VR}{virtual reality}
 \newacro{UL}{uplink}
 \newacro{OFDM}{orthogonal frequency division multiplexing}
 \newacro{RF}{radio frequency}
 \newacro{TX}{transmitter}
 \newacro{RX}{receiver}
 \newacro{FFT}{fast Fourier transform}
 \newacro{IFFT}{inverse fast Fourier transform}
 \newacro{DFT}{discrete Fourier transform}
 \newacro{IDFT}{inverse discrete Fourier transform}
 \newacro{QPSK}{quadrature phase shift keying}
 \newacro{QAM}{quadrature amplitude modulation}
 \newacro{M}{modulation order}
 \newacro{EVM}{error vector magnitude}
 \newacro{DAC}{digital-to-analog converter}
 \newacro{CP}{cyclic prefix}
 \newacro{ISI}{inter-symbol interference}
 \newacro{FIR}{finite impulse response}
 \newacro{CIR}{channel impulse response}
 \newacro{LS}{least square}
 \newacro{MMSE}{minimum mean square error}
 \newacro{MSE}{mean square error}
 \newacro{CSI}{channel state information}
 \newacro{DPA}{data pilot-aided}
 \newacro{DL}{deep learning}
 \newacro{FCNN}{fully-connected neural network}
 \newacro{CNN}{convolutional neural network}
 \newacro{NN}{neural network}
 \newacro{FW-P}{forward propagation}
 \newacro{BW-P}{backward propagation}
 \newacro{FC}{fully-connected}
 \newacro{DNN}{deep neural network}
 \newacro{Uni-DNN}{universal deep neural network}
 \newacro{BS}{base station}
 \newacro{UE}{user equipment}
 \newacro{BER}{bit error rate}
 \newacro{SNR}{signal-to-noise ratio}
 \newacro{SER}{symbol error rate}
 \newacro{CO2}{carbon dioxide}
 \newacro{DDCE}{decision directed channel estimation}
 \newacro{RE}{resource element}
 \newacro{PAPR}{peak-to-average power ratio}
 \newacro{CFR}{channel frequency response}
 \newacro{SVD}{singular value decomposition}
 \newacro{SBS}{symbol-by-symbol}
 \newacro{ZF}{zero forcing}
 \newacro{FBF}{frame-by-frame}
 \newacro{AWGN}{additive white Gaussian channel noise}
 \newacro{ML}{maximum likelihood}
 \newacro{SISO}{single-input-single-output}
 \newacro{E2E}{end-to-end}
 \newacro{i.i.d}{independent identically distributed}
 \newacro{QoS}{quality of service}
 \newacro{LOS}{line of sight}
 \newacro{NLOS}{non line of sight}
 \newacro{RGB}{red, green and blue}
 \newacro{PDF}{probability density function}
 \newacro{CDL}{clustered delay line}
 \newacro{TDL}{tapped delay line}
 \newacro{WINNER II}{Wireless World Initiative New Radio II}
 \newacro{MIMO}{multiple-input-multiple-output}
 \newacro{RMS}{root mean square}
 \newacro{FD}{fractional delay}
 \newacro{CORDIS}{Community Research and Development Information Service}
 \newacro{MRC}{maximum ratio combining}
 \newacro{MAP}{maximum a posteriori}
 \newacro{SVD}{singular value decomposition}
 \newacro{FDE}{frequency domain channel estimation and equalization}
 \newacro{CE}{categorical cross-entropy}
 
 \acrodef{LTI}{Linear time-invariant}
 \acrodef{3GPP}{Third Generation Partnership Project}
 \acrodef{2G}{second-generation}
 \acrodef{GSM}{global system for mobile communications}
 \acrodef{EDGE}{enhanced data rates for global evolution}
 \acrodef{IS-95}{interim standard 95}
 \acrodef{3G}{third-generation}
 \acrodef{4G}{fourth-generation}
 \acrodef{5G}{fifth-generation}
 \acrodef{mmWave}{millimeter wave}
 \acrodef{6G}{sixth-generation}
 \acrodef{LTE}{long-term evolution}
 \acrodef{PSS}{primary synchronization signal}
 \acrodef{SSS}{secondary synchronization signal}
 \acrodef{TDMA}{time-division-multiple-access}
 \acrodef{FDMA}{frequency division multiple access}
 \acrodef{SC-FDMA}{single carrier frequency division multiple access}
 \acrodef{CDMA}{code division multiple access}
 \acrodef{PN-codes}{pseudo random codes}
 \acrodef{VR}{virtual reality}
 \acrodef{UL}{uplink}
 \acrodef{OFDM}{orthogonal frequency division multiplexing}
 \acrodef{RF}{radio frequency}
 \acrodef{TX}{transmitter}
 \acrodef{RX}{receiver}
 \acrodef{FFT}{fast Fourier transform}
 \acrodef{IFFT}{inverse fast Fourier transform}
 \acrodef{DFT}{discrete Fourier transform}
 \acrodef{IDFT}{inverse discrete Fourier transform}
 \acrodef{QPSK}{quadrature phase shift keying}
 \acrodef{QAM}{quadrature amplitude modulation}
 \acrodef{M}{modulation order}
 \acrodef{EVM}{error vector magnitude}
 \acrodef{DAC}{digital-to-analog converter}
 \acrodef{CP}{cyclic prefix}
 \acrodef{ISI}{inter-symbol interference}
 \acrodef{FIR}{finite impulse response}
 \acrodef{CIR}{channel impulse response}
 \acrodef{LS}{least square}
 \acrodef{MMSE}{minimum mean square error}
 \acrodef{MSE}{mean square error}
 \acrodef{CSI}{channel state information}
 \acrodef{DPA}{data pilot-aided}
 \acrodef{DL}{deep learning}
 \acrodef{FCNN}{fully-connected neural network}
 \acrodef{CNN}{convolutional neural network}
 \acrodef{NN}{neural network}
 \acrodef{FW-P}{forward propagation}
 \acrodef{BW-P}{backward propagation}
 \acrodef{FC}{fully-connected}
 \acrodef{DNN}{deep neural network}
 \acrodef{Uni-DNN}{universal deep neural network}
 \acrodef{BS}{base station}
 \acrodef{UE}{user equipment}
 \acrodef{BER}{bit error rate}
 \acrodef{SNR}{signal-to-noise ratio}
 \acrodef{SER}{symbol error rate}
 \acrodef{CO2}{carbon dioxide}
 \acrodef{DDCE}{decision directed channel estimation}
 \acrodef{RE}{resource element}
 \acrodef{PAPR}{peak-to-average power ratio}
 \acrodef{CFR}{channel frequency response}
 \acrodef{SVD}{singular value decomposition}
 \acrodef{SBS}{symbol-by-symbol}
 \acrodef{ZF}{zero forcing}
 \acrodef{FBF}{frame-by-frame}
 \acrodef{AWGN}{additive white complex Gaussian channel Noise}
 \acrodef{ML}{maximum likelihood}
 \acrodef{SISO}{single-input-single-output}
 \acrodef{E2E}{end-to-end}
 \acrodef{i.i.d}{independent identically distributed}
 \acrodef{QOS}{quality of service}
 \acrodef{LOS}{line of sight}
 \acrodef{NLOS}{non line of sight}
 \acrodef{RGB}{red, green and blue}
 \acrodef{PDF}{probability density function}
 \acrodef{CDL}{clustered delay line}
 \acrodef{TDL}{tapped delay line}
 \acrodef{WINNER II}{Wireless World Initiative New Radio II}
 \acrodef{MIMO}{multiple-input-multiple-output}
 \acrodef{RMS}{root mean square}
 \acrodef{FD}{fractional delay}
 \acrodef{CORDIS}{Community Research and Development Information Service}
 \acrodef{MRC}{maximum ratio combining}
 \acrodef{MAP}{maximum a posteriori}
 \acrodef{SVD}{singular value decomposition}
 \acrodef{FDE}{frequency domain channel estimation and equalization}
 \acrodef{CE}{categorical cross-entropy}

\title{A Universal Deep Neural Network for Signal Detection in Wireless Communication Systems}
\author{Khalid Albagami$^1$, Nguyen Van Huynh$^2$, and Geoffrey Ye Li$^1$\\
	$^1$ Department of Electrical and Electronic Engineering, Imperial College London, UK\\
	$^2$ School of Computing, Engineering and the Built Environment, Edinburgh Napier University, UK\\
	Emails: khalid.albagami22@imperial.ac.uk, h.nguyen2@napier.ac.uk, geoffrey.li@imperial.ac.uk}
\maketitle
\begin{abstract}




Recently, deep learning (DL) has been emerging as a promising approach for channel estimation and signal detection in wireless communications. The majority of the existing studies investigating the use of DL techniques in this domain focus on analysing channel impulse responses that are generated from only one channel distribution such as additive white Gaussian channel noise and Rayleigh channels. In practice, to cope with the dynamic nature of the wireless channel, DL methods must be re-trained on newly non-aged collected data which is costly, inefficient, and impractical. To tackle this challenge, this paper proposes a novel universal deep neural network (Uni-DNN) that can achieve high detection performance in various wireless environments without retraining the model. In particular, our proposed Uni-DNN model consists of a wireless channel classifier and a signal detector which are constructed by using DNNs. The wireless channel classifier enables the signal detector to generalise and perform optimally for multiple wireless channel distributions. In addition, to further improve the signal detection performance of the proposed model, convolutional neural network is employed. Extensive simulations using the orthogonal frequency division multiplexing scheme demonstrate that the bit error rate performance of our proposed solution can outperform conventional DL-based approaches as well as least square and minimum mean square error channel estimators in practical low pilot density scenarios.
\end{abstract}


\section{Introduction}
\label{Sec:intro}


Channel estimation and signal detection's role in wireless communication is to accurately estimate the characteristics of the wireless channel and enable data recovery with low error rates while maintaining acceptable spectral efficiency and overall network performance. The role is essential, especially in scenarios where the wireless channel is highly dynamic and subject to variations due to factors like mobility and interference \cite{YonJaeWonChu:2010}. The deployed data pilot-aided methods in practice are sub-optimal for highly frequency-selective wireless channels. In addition, conventional data pilot-aided methods like \ac{LS}, suffer from interference and noise amplification at low \ac{SNR} scenarios while approaches like \ac{MMSE}, which leverage second-order channel statistics to improve the \ac{LS} solution, demand significant computational resources \cite{OzdMehArsHus:2007}, \cite{VanOveMagSarPer:1995}.



To address these drawbacks, data-driven approaches such as \ac{DL} that utilises \ac{DNN} can improve channel estimation accuracy, reduce pilot overhead, and enhance the resilience of wireless communication systems in dynamic and challenging environments without the requirement of prior channel knowledge. 
In \cite{YeHaoGeoJuaBii:2018}, a 3-hidden-layer fully connected neural network jointly estimates the channel and detects the signal, achieving performance comparable to the \ac{MMSE} estimator and demonstrating robustness across various test conditions. However, most of the previous research in this field that explores the application of \ac{DL} methods primarily concentrates on the analysis of channel impulse responses produced by single channel distributions like \ac{AWGN} and Rayleigh channels. In real-world scenarios, adapting \ac{DL} models to the ever-changing wireless channel conditions requires periodic retraining on freshly collected data, which is not only expensive but also inefficient and impractical. In addition to other challenges such as the difficulty of data collection, lack of generalisation in real noise and interference, interpretability and theoretical guarantees and energy inefficiency \cite{YeHaoGeoJuaBii:2018}, \cite{YanYuwGaoFeiMaZha:2019}, \cite{MaYeLi:2018}.


To address the non-generalization shortcoming, our article aims to develop a novel \ac{Uni-DNN} model consisting of a wireless channel classifier and a signal detector which are constructed by using \ac{DNN}s. The wireless channel classifier is used to detect the wireless channel type and then its output is fed to the signal detector along with the received signal. This allows the proposed model to achieve high detection performance in various wireless environments without the need to retrain the model. Furthermore, this can reduce the dependency on pilots and reduce the computation power required to deploy \ac{DL} models in practice. Also, this would enable existing wireless networks to function at a lower \ac{BER} for a specific \ac{SNR}, thereby improving the overall network coverage.




\section{System Model}
\label{Sec.System}

\begin{figure*}[ht]
\centering
\includegraphics[width=2\columnwidth]{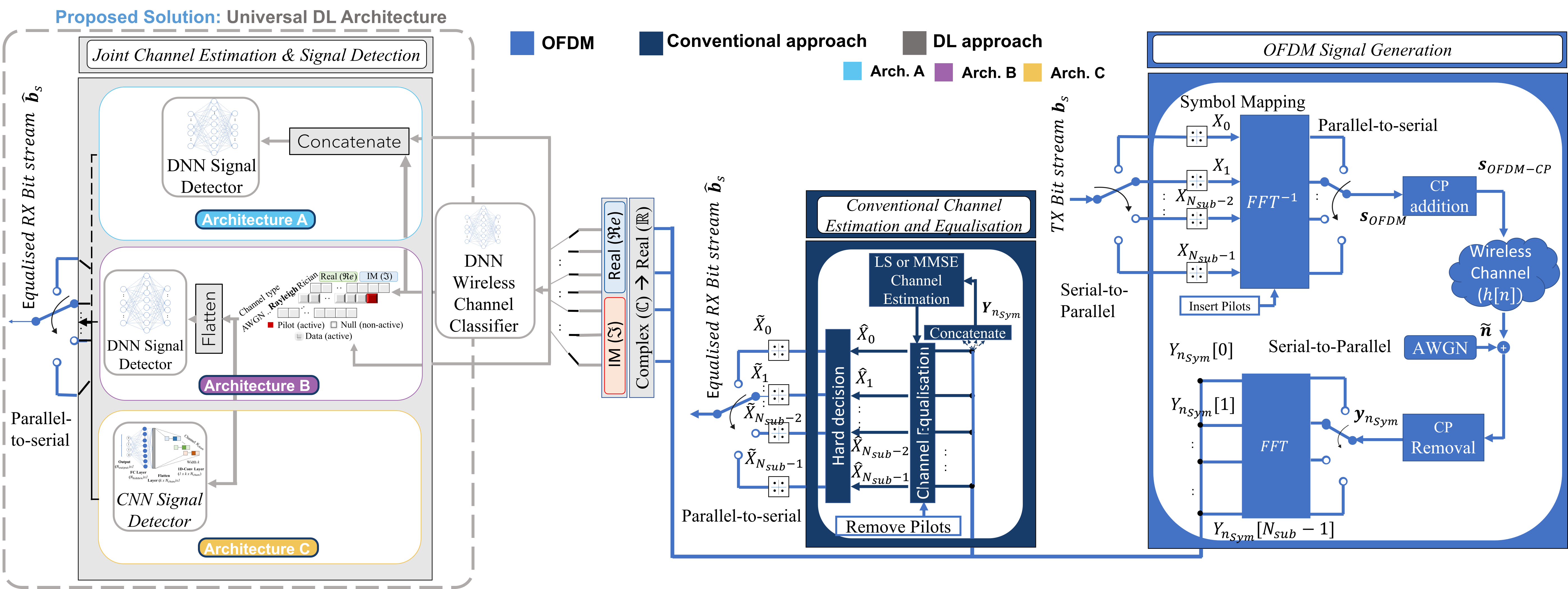}
\caption[System model consisting of OFDM signal generation, conventional and joint channel estimation and signal detection.]{System model consisting of OFDM signal generation, conventional and joint channel estimation and signal detection.}
\label{fig_1.1}
\end{figure*}

In this work, we consider a typical \ac{SISO} \ac{OFDM} system as illustrated as part of Fig.~\ref{fig_1.1}, where three blocks are discussed namely \ac{OFDM} signal generation, conventional and joint channel estimation and signal detection. In the \ac{OFDM} signal generation block, $\boldsymbol{b}_s$ is the transmitted bit stream, $\hat{\boldsymbol{b}}_s$ is the received equalised bit stream, $\boldsymbol{s}_\mathrm{OFDM}$ is a $N_\mathrm{IFFT}$-long vector of one \ac{OFDM} time domain sampled symbol, $\boldsymbol{s}_\mathrm{OFDM-CP}$ is a $(N_\mathrm{IFFT}+N_\mathrm{CP})$-long vector of an \ac{OFDM} symbol after adding the \ac{CP}, $\boldsymbol{y}_\mathrm{n_\mathrm{Sym}}$ is a $(N_\mathrm{IFFT}+N_\mathrm{taps}-1)$-long vector of the $n_\mathrm{Sym}^\mathrm{th}$-received \ac{OFDM} symbol after removing the CP and $\hat{X}_0, ..., \hat{X}_\mathrm{N_\mathrm{sub}-1}$ are the received \ac{OFDM} sub-carriers after equalisation where $N_\mathrm{sub}$ is the number of \ac{OFDM} sub-carriers. Without loss of generality in this article we assume the use of gray-coded \ac{QPSK} for $X_0, ..., X_\mathrm{N_\mathrm{sub}-1}$. Also, for pilot insertion within the \ac{OFDM} resource grid we assume comb-type pilot architecture. The transmitted \ac{OFDM} sub-carriers can be expressed as follows,

\begin{equation}
\label{OFDM_received_sig_eq1}
\boldsymbol{X}=\left[\begin{array}{cccc}
X_0 & 0 & \ldots & 0 \\
0 & X_1 & & \vdots \\
\vdots & & \ddots & 0 \\
0 & \ldots & 0 & X_{N_{s u b}-1}
\end{array}\right], \\ 
\end{equation}
where $\boldsymbol{X}$ is a $N_\mathrm{s u b} \times N_\mathrm{s u b}$ diagonal matrix storing the transmitted digitally modulated $N_\mathrm{s u b}$ \ac{OFDM} sub-carriers. Then, the transmitted signal is passed through the \ac{CIR} which can be expressed as,
\begin{equation}
\label{OFDM_received_sig_eq2}
\boldsymbol{H}=\left[\begin{array}{c}
H_0 \\
H_1 \\
\vdots \\
H_{N_{s u b}-1}
\end{array}\right], \\
\end{equation}
where $\boldsymbol{H}$ is a $N_\mathrm{s u b}$-column vector storing the frequency response of \ac{CIR} at each \ac{OFDM} sub-carrier. After that, we incorporate the \ac{AWGN} effect that can be expressed as, 
\begin{equation}
\label{OFDM_received_sig_eq3}
\tilde{\boldsymbol{n}}=\left[\begin{array}{c}
\tilde{n}_0 \\
\tilde{n}_1 \\
\vdots \\
\tilde{n}_{N_{s u b}-1}
\end{array}\right], \\
\end{equation}
where $\tilde{\boldsymbol{n}}$ is a $N_{s u b}$-column vector containing the \ac{AWGN} experienced by the transmitted \ac{OFDM} sub-carriers. The received \ac{OFDM} sub-carriers then can be expressed as follows,
\begin{equation}
\label{OFDM_received_sig_eq4}
\boldsymbol{Y}_{n_{\text {sym }}}=\boldsymbol{X} \boldsymbol{H}+\boldsymbol{\tilde{n}},
\end{equation}
where the $N_\mathrm{s u b}$-column vector storing the received \ac{OFDM} sub-carriers $\boldsymbol{Y}_\mathrm{n_\mathrm{Sym}}$ is the result of the matrix multiplication of $\boldsymbol{X}$ and $\boldsymbol{H}$ added to $\boldsymbol{\tilde{n}}$. 

In this study, we employ true channel performance along with the widely adopted \ac{LS} and \ac{MMSE} \ac{OFDM} channel estimators as benchmarks to assess the performance of \ac{DL}-based joint channel estimation and signal detection. The \ac{LS} estimator expressed in \eqref{LS_CH_estimator_final} minimizes the \ac{MSE} of $\|\boldsymbol{Y}_{n_{Sym}}-\boldsymbol{X} \hat{\boldsymbol{H}}\|^2$ \cite{YonJaeWonChu:2010}.
\begin{equation}
\label{LS_CH_estimator_final}
\hat{\boldsymbol{H}}_{L S}=\boldsymbol{X}^{-1} \boldsymbol{Y}_{n_{Sym}}.
\end{equation}
Whereas, the \ac{MMSE} channel estimator minimizes the error expression $E\left\{\|\boldsymbol{H}-\boldsymbol{W}\hat{\boldsymbol{H}}_{L S}\|^2\right\}$ where $\boldsymbol{W}$ is a weight matrix \cite{YonJaeWonChu:2010}. The final perfect-\ac{CSI} and non-perfect-\ac{CSI} \ac{MMSE} expressions are shown in \eqref{MMSE_CH_estimator_perfectCSI} and \eqref{MMSE_CH_estimator_nonperfectCSI}, respectively.
\begin{equation}
\label{MMSE_CH_estimator_perfectCSI}
\hat{\boldsymbol{H}}_{\mathrm{M M S E_{perfectCSI}}}=\boldsymbol{R}_{H H}\left(\boldsymbol{R}_{H H}+\frac{\sigma_n}{\sigma_X} \boldsymbol{I}\right)^{-1} \hat{\boldsymbol{H}}_{L S},
\end{equation}
\begin{multline}
\label{MMSE_CH_estimator_nonperfectCSI}
\hat{\boldsymbol{H}}_{M M S E_{nonperfectCSI}}= \\ \boldsymbol{R}_{\hat{H}_{LS} \hat{H}_{LS}}\left(\boldsymbol{R}_{\hat{H}_{LS} \hat{H}_{LS}}+\frac{\sigma_n}{\sigma_X} \boldsymbol{I}\right)^{-1} \hat{\boldsymbol{H}}_{L S}.
\end{multline}
In \eqref{MMSE_CH_estimator_perfectCSI} and \eqref{MMSE_CH_estimator_nonperfectCSI}, $\boldsymbol{R}_{H H}$ and $\boldsymbol{R}_{\hat{H}_{LS} \hat{H}_{LS}}$ are the auto-correlation of the true and \ac{LS} estimated channel, respectively and $\frac{\sigma_n}{\sigma_X}$ is the reciprocal of the \ac{SNR}. Now to estimate channels at non-pilot OFDM sub-carriers, we use linear interpolation and for data sub-carriers that are outside of the interpolation range, various techniques exist, like virtual sub-carrier insertion. For simplicity, we assign the nearest pilot sub-carrier's estimate to the out-of-boundary data sub-carriers. After channel estimation, we apply maximum likelihood signal detection for \ac{OFDM} \ac{SISO}, to recover $\boldsymbol{X}$ as follows,
\begin{equation}
\label{ML_sig_detector_eq}
\begin{gathered}
D(\boldsymbol{H}, \boldsymbol{X})=\|\boldsymbol{Y}_{n_{Sym}}-\boldsymbol{X} \boldsymbol{H}\|^2, \\
diag(\hat{\boldsymbol{X}})=\underset{\boldsymbol{X}=\hat{\boldsymbol{X}}}{\operatorname{argmin }} \ D(\boldsymbol{H}, \boldsymbol{X})=\boldsymbol{Y}_{n_{\text {sym }}} / \hat{\boldsymbol{H}},
\end{gathered}
\end{equation}
where $D(\boldsymbol{H}, \boldsymbol{X})$ is the distance function and $diag(.)$ is the matrix diagonal symbol  \cite{PeiKob:2002}. Finally, the result from \eqref{ML_sig_detector_eq} undergoes a hard decision process and symbol demapping to obtain $\hat{\boldsymbol{b}}_s$.

\section{Deep Learning (DL) based Channel Estimation and Signal Detection}
\label{Deep learning}


\subsection{Conventional \ac{DL}-based approach}
\label{coventional Deep learning}


In non-complex \ac{DL} joint channel estimation and signal detection, $N_\mathrm{sub}$ received complex symbols $\boldsymbol{Y}_\mathrm{n_\mathrm{Sym}}$ are split to $2N_\mathrm{sub}$ concatenated real and imaginary values before being fed as input to the \ac{DL} model. The output of the model $\hat{\boldsymbol{y}}_i$ is the recovered bits. The sigmoid activation function is employed at the output layer, yielding an output in the range of 0 to 1 suitable for the application. It can be expressed as follows,
\begin{equation}
\label{sigmoid_der_eq}
\begin{gathered}
\phi\left(\boldsymbol{z}_i^{(l)}\right)=\frac{1}{1+e^{-\boldsymbol{z}_i^{(l)}}}, \\
\end{gathered}
\end{equation}
where $\boldsymbol{z}_i^{(l)}$ is the output vector of the $i^{th}$ sample in \ac{NN} layer $l$, and $\phi(.)$ denotes a non-linear activation function. For hidden layers, the ReLU activation function is chosen for its constant gradient properties, which mitigate the risk of encountering vanishing gradient problem. It can be expressed as follows,
\begin{equation}
\label{Relu_der_eq}
\begin{gathered}
\boldsymbol{\phi}\left(\boldsymbol{z}_i^{(l)}\right)=\max _{\text {row-by-row }}\left(\boldsymbol{0}_{N_{\text {hidden }}+\boldsymbol{1}}, \boldsymbol{z}_i^{(l)}\right), \\
\end{gathered}
\end{equation}
where $\boldsymbol{0}_{N_{\text {hidden }}+\boldsymbol{1}}$ represents a $N_{\text {hidden }}$-column vector of zero elements and $\max _{\text {row-by-row }}$ represents a row-by-row nonlinear maximizing function. Furthermore, the ReLU sparsity feature, which avoids activating neurons with negative input, accelerates convergence and lowers computational complexity. \ac{MSE} loss function $\boldsymbol{J}(w)$ is used to guide the \ac{DL} model toward optimal performance, which is optimal in the presence of complex Gaussian distributed channels and noise. It also serves as a suitable metric for assessing the proximity of the estimated bit stream to the transmitted bit stream. The \ac{MSE} loss function can be expressed as follows,
\begin{equation}
\label{MSE_loss_func_eq}
\boldsymbol{J}_{MSE}(w)=\frac{1}{2} \sum\left(\boldsymbol{y}_i-\hat{\boldsymbol{y}}_i\right)^2,  
\end{equation}
where $\boldsymbol{J}_{MSE}(w)$, $\boldsymbol{y}_i$ and $\hat{\boldsymbol{y}}_i$ represent the \ac{MSE} loss function, the transmitted bits, and the \ac{DL} model predicted bits, respectively.


A drawback of using \ac{DL} model in the application of joint channel estimation and signal detection is that pre-training a \ac{DL} model on a dataset generated from one particular channel model may yield sub-optimal performance when tested on a different channel model. In addition to other challenges such as the difficulty of data collection, lack of generalisation in real noise and interference, interpretability and theoretical guarantees and energy inefficiency. To enhance the conventional \ac{DL} model performance, one solution is to train it on a dataset containing multiple channel models. However, training the \ac{DL} model on a dataset comprising various correlated non-\ac{i.i.d} channel models constrains the achievable overall \ac{MSE} floor.


\subsection{Universal Deep Neural Network (Uni-DNN) for Channel Estimation and Signal Detection}
\label{Uni-DNN section}



To tackle the sub-optimality of the conventional \ac{DL} approach, we propose a novel \ac{DL} architecture comprising two cascaded \ac{NN}s. The first \ac{NN} serves as a wireless channel classifier taking the concatenated real and imaginary parts of $\boldsymbol{Y}_\mathrm{n_\mathrm{Sym}}$ as input and inferring the correct channel class using one-hot encoding. The second \ac{NN} model works as a signal detector to recover $\boldsymbol{b}_s$, sharing the same input as the first \ac{NN} but incorporating additional information about the channel class inferred by the first \ac{NN}. This extra information has the potential to enhance the multi-channel \ac{DL} model and narrow the performance gap with the single-channel trained \ac{DL} model. Moreover, the classifier is not limited to channel type classification; it can be extended to various classifications that offer valuable insights for improving the signal detector \ac{NN}'s performance. Examples of such channel-related information include the number of taps, delay spread, and Doppler spread estimations of the channel model. In this way, the proposed \ac{Uni-DNN} can be generalized for different channels and settings without relying on specific channel parameters for good detection performance. To construct the channel classifier \ac{DNN} dataset, practical steps involve gathering data from the wireless environment, followed by using unsupervised learning to cluster it into various channel models. Subsequently, offline training is conducted to optimize the \ac{Uni-DNN} architecture. Once the model meets performance requirements, such as a specific \ac{BER} or quality of service, it can be deployed online for real-time data inferences. Collecting a diverse dataset from various wireless channels across the globe is important to achieve consistent results.

With the clustered dataset, the proposed universal \ac{DL} architecture A employs 2-cascaded \ac{DNN} as depicted in Fig.~\ref{fig_1.1}. where the channel classifier \ac{DNN} is trained on a dataset where the input is the concatenated real and imaginary parts of $\boldsymbol{Y}_\mathrm{n_\mathrm{Sym}}$, and the output is the one-hot encoded class of the wireless channel model. Once the channel classifier \ac{DNN} is trained, the optimal \ac{DNN} layers' weights are saved. Then, the second \ac{DNN} is trained on the same input concatenated with the first \ac{DNN} predicted channel one-hot encoded class. Training the signal detector \ac{DNN} on the classifier's predictions rather than the ground-truth channel classes provides better overall generalisation. Once the second training cycle is done for the signal detector \ac{DNN}, \ac{Uni-DNN} architecture A is ready to make inferences.

While the \ac{Uni-DNN} architecture A has the potential to enhance multi-channel \ac{DNN}s, it has drawbacks. In particular, the system's computational complexity can be increased compared to multi-channel \ac{DNN}s, as it employs two cascaded \ac{DNN}s. Another issue is that the overall model's performance is affected not only by signal detector errors but also by misclassifications made by the channel classifier \ac{DNN}. Furthermore, the enhancement obtained by the proposed \ac{Uni-DNN} architecture relies on how effectively we incorporate additional channel information into the signal detector \ac{DNN}. Two other \ac{Uni-DNN} architectures, labelled B and C as in Fig.~\ref{fig_1.1} , are also analyzed. In architecture B, a 2D-grid representation is utilised where the rows represent the real and imaginary parts of $\boldsymbol{Y}_\mathrm{n_\mathrm{Sym}}$ while columns represent different channel model types, with only one activated channel model at a time based on the channel classifier \ac{DNN} output. Then, the 2D-grid tensor is flattened and fed to the \ac{FC} layer. In architecture C, instead of flattening, \ac{CNN} is employed to extract the important information from the 2D-grid. As illustrated in Fig.~\ref{fig_1.1}, the 2D-grid is processed differently, resembling a 1D-vector with $N_{chan}$ channels, akin to an image with RGB channels. The latter two architectures enable the \ac{FC} layer to assign distinct weights to each wireless channel model, rather than using an overall weight aimed at minimizing the overall \ac{MSE} error floor. This, combined with the use of a \ac{CNN} model adept at analyzing image-like datasets, has the potential to improve \ac{MSE} performance and speed up model convergence. However, a drawback of architectures B and C is the increased complexity, as the input size is expanded by a factor of $N_{chan}$.


\subsection{Analytical Computational Complexity Comparison}
\label{Analytical complexity comparison}


In this section, we will delve into the computational complexity of both conventional and \ac{DL} models for channel estimation and signal detection at the receiver, focusing on the number of multiplications and relative operation time. Table \ref{infernce_complexity} shows the analytical inference time complexities. Since $N_{out} \approx N_{in}$ in low pilot frequency settings, we can assume $\mathcal{O}(N_{in}) = \mathcal{O}(2N_{sub}) \approx \mathcal{O}(N_{sub})$ in case of all \ac{DL} models except for \ac{Uni-DNN} architecture B and C where $\mathcal{O}(N_{in}) = \mathcal{O}(N_{chan} \times N_{sub})$ to simplify the comparison. Table \ref{infernce_complexity} demonstrates that analytical results place \ac{DL} models between \ac{MMSE} and \ac{LS} methods in terms of complexity, offering substantial inference time reductions compared to \ac{MMSE}.

\begin{table}[ht]
\centering
\caption{Model inference analytical complexity analysis.}
\label{infernce_complexity}
\begin{tabular}[t]{cc}
\hline
Model/Parameter & Inference complexity \\
\hline
\ac{LS} & $\mathcal{O}(N_{sub})$ \\
\ac{MMSE} & $\mathcal{O}(N_{sub}^{2} \times m)$\\
Single-channel & $\mathcal{O}(N_{hid} \times N_{sub})$\\
Multi-channel & $\mathcal{O}(N_{hid} \times N_{sub})$\\
Uni-Arc-A & $2 \times \mathcal{O}(N_{hid} \times N_{sub})$ \\
Uni-Arc-B & $\mathcal{O}(N_{hid} \times N_{sub}) + \mathcal{O}(N_{hid} \times N_{sub} \times N_{chan})$ \\
Uni-Arc-C & $\mathcal{O}(N_{hid} \times N_{sub} +N_{chan}^2 \times k \times l \times N_{sub})$ \\
\hline
\end{tabular}
\end{table}%


\section{Simulation Setup and Results}

\subsection{Wireless Channel Models Generation}
\label{channel_model}

In this study, the \ac{OFDM} generated transmitted symbols are passed through various simulated frequency selective and frequency flat fast fading wireless channel models namely, Rayleigh, Rician, 3GPP \ac{TDL}-A, WINNER II and \ac{AWGN}-only impaired channels. Rayleigh channel is generated by utilizing the Monte-Carlo approach to simulate the magnitude of a complex number consisting of \ac{i.i.d} normally distributed random variable with zero mean and unit variance real and imaginary parts. Rician channel is simulated by combining a Rayleigh distributed \ac{NLOS} component with \ac{LOS} component of fixed magnitude and uniformly distributed phase between $\left[-\frac{\pi}{2}, \frac{\pi}{2}\right]$. The contribution of each component is determined by the kappa factor $\kappa$ where in the simulation $\kappa=2$, i.e, \ac{LOS} component is twice as strong as the \ac{NLOS} component \cite{LuoZhoZhaYanJonEdm2020}. 3GPP \ac{TDL}-A is a \ac{NLOS} channel model based on practical measurements for frequencies from 0.5 GHz to 100 GHz done by 3GPP Radio Access Network Technical Specification Group \cite{3GPP:2017}. The steps adopted to generate 3GPP \ac{TDL}-A channel are depicted in \cite{3GPP:2017} where the model is flexible to simulate a flat-fading to a highly selective scenario. Finally, WINNER II channel model is based on the WINNER II initiative led by Community Research and Development Information Service to develop a radio access network system that can simulate many radio environment scenarios for short and long range \cite{YeHaoGeoJuaBii:2018}, \cite{Kyö:2008}. The simulation parameters for \ac{OFDM}, \ac{DL} models and the channel models are outlined in Table \ref{OFDM simulation parameters table}, \ref{Channel models simulation parameters table} and \ref{Deep Learning model simulation parameters table} respectively, where $L$ and $L_{frac}$ are the number of channel and fractional taps \cite{3GPP:2017}, \cite{Kyö:2008} respectively, $\sigma_{DS}$ is the delay spread, and $m_{H}$ is the number of channel samples to generate $\boldsymbol{R}_{H H}$ and $\boldsymbol{R}_{\hat{H}_{LS} \hat{H}_{LS}}$in Monte-Carlo simulations.


\begin{table}[ht]
\centering
\caption{OFDM simulation parameters.}
\label{OFDM simulation parameters table}
\begin{tabular}[t]{cc}
\hline
Parameter &Value\\
\hline
Number of bits ($N_{bits}$) & $10^7$\\
Number of pilots ($N_p$) & 8, 16, 32\\
Pilot frequency ($p_f$) & 8, 4, 2\\
\ac{FFT} window length ($N_{FFT}$) & 64\\
Inverse \ac{FFT} window length ($N_{IFFT}$) & 64\\
Number of sub-carriers ($N_{sub}$) & 64\\
Sub-carrier spacing ($\Delta f$) & 15kHz\\
\ac{OFDM} symbol period ($T_{Sym}$) & 66.67 $\mu$s\\
\ac{OFDM} sampling frequency ($f_s$) & 0.96 MHz\\
\ac{OFDM} sampling period ($T_{s}$) & 1.04 $\mu$s\\
Cyclic prefix (CP) length ($N_{CP}$) & 16 samples\\
Cyclic prefix (CP) period ($T_{CP}$) & 16.67 $\mu$s\\
Noise source ($\tilde{\boldsymbol{n}}$) & \ac{AWGN}\\
\hline
\end{tabular}
\end{table}%

\begin{table}[ht]
\centering
\caption{Deep learning models' simulation parameters.}
\label{Deep Learning model simulation parameters table}
\begin{tabular}[t]{cc}
\hline
Parameter &Value\\
\hline
Number of input features ($N_{in}$) & $128$\\
Number of hidden layers & $1$\\
Number of neurons ($N_{hid}$) & $256-1024$\\
\ac{CNN} 1D kernel dim. ($N_{chan} \times k$) & $5 \times 1$\\
\ac{CNN} number of filters ($n$) & $5$\\
\ac{CNN} type of padding & $Same$\\
Number of output features ($N_{out}$) & $64, 96, 112$\\
Number of samples ($m$) & $100,000$\\
Training/Validation split & $70\% \ / \ 30\%$\\
Number of epochs ($N_{epochs}$) & 700\\
Batch size ($N_{batch}$) & $3,000 \ samples$\\
Activation function & ReLU (hidden) - Sigmoid (output)\\
Optimizer & ADAM\\
Learning rate ($\alpha$) & 0.001\\
L2-Regularization & $2 \times 10^{-6}$\\
Dropout Rate & 0.01\\
Performance Metrics & Binary accuracy, \ac{MSE}, \ac{BER}\\
\hline
\end{tabular}
\end{table}%

\begin{table}[ht]
\centering
\caption{Channel models simulation parameters.}
\label{Channel models simulation parameters table}
\small
\begin{tabular}[t]{ccccc}
\hline
Param./Chan. & Rayleigh & Rician & \ac{TDL}-A & WINNERII\\
\hline
$L$ & $4$ & $6$ & $~1$ & $~5$\\
$L_{frac}$ & - & - & $23$ & $24$\\
$\sigma_{DS}$ & $4.160 \mu s$ & $6.240 \mu s$ & $0.965 \mu s$ & $5.200 \mu s$\\
Freq. Selectivity & selective &  selective &  flat & selective\\
Fading rate & fast & fast &  fast & fast\\
$m_{H}$ & 1,000 & 1,000 & 1,000 & 1,000\\
\hline
\end{tabular}
\end{table}%

\subsection{Bit Error Rate Simulation Results}
\label{DL model BER}



In this section, we analyze and compare the performance of the various \ac{DL} models proposed in this article, evaluating their \ac{BER} across \ac{SNR} values from 0 dB to 20 dB and their image detection performance at 20 dB. We benchmark the \ac{DL} models against ideal true channel performance and conventional methods, specifically \ac{LS} and \ac{MMSE}. Due to space constraints, we present results for only the 3GPP \ac{TDL}-A and Rician channel models at $N_p=8$ and $N_p=32$. The other channel models in Table \ref{Channel models simulation parameters table} are showing similar results.

\begin{figure}[ht]
\centering
\includegraphics[width=0.79\columnwidth]{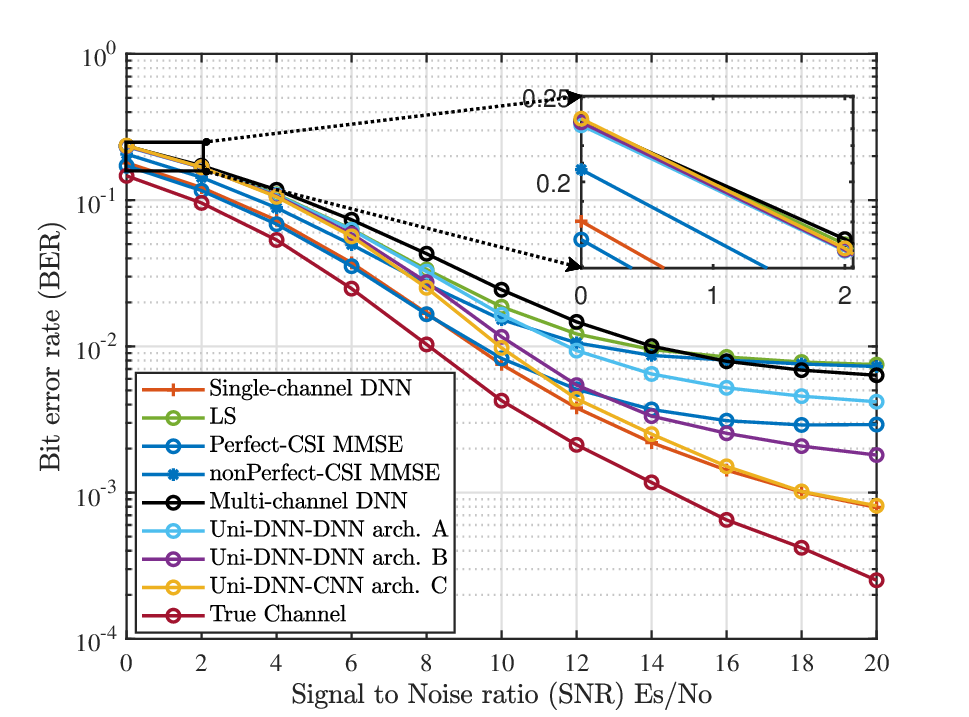}
\caption[8-pilots 3GPP TDL-A channel BER performance vs. SNR for conventional and DL methods comparison.]{8-pilots 3GPP TDL-A channel BER performance vs. SNR for conventional and DL methods comparison.}
\label{fig_1.2}
\end{figure}

\begin{figure}[ht]
\centering
\includegraphics[width=\columnwidth]{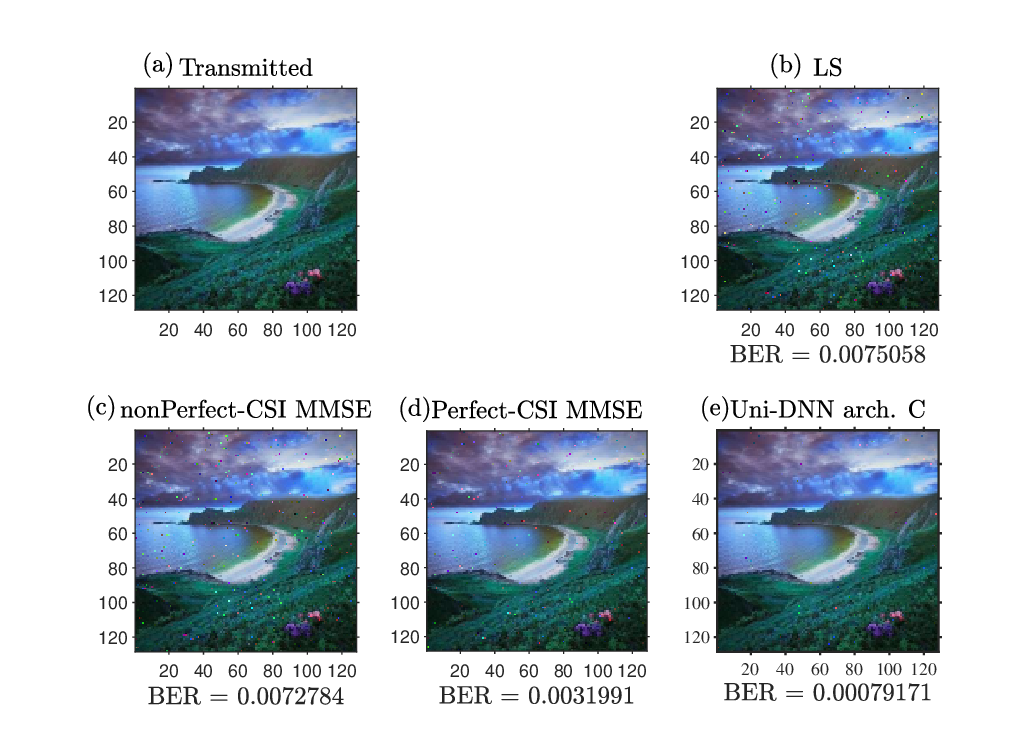}
\caption[3GPP TDL-A channel (a) Transmitted, (b) LS, (c) non-perfect, (d) perfect MMSE and (e) DNN equalised image at 20 dB.]{3GPP TDL-A channel (a) Transmitted, (b) LS, (c) non-perfect, (d) perfect MMSE and (e) DNN equalised image at 20 dB.}
\label{fig_1.3}
\end{figure}

Fig.~\ref{fig_1.2} shows that employing \ac{Uni-DNN} architectures A, B and C outperforms both conventional and multi-channel \ac{DL} model and closes the gap in \ac{BER} performance to the single-channel trained \ac{DL} model for 3GPP \ac{TDL}-A channel at high \ac{SNR}. For instance, Fig.~\ref{fig_1.3} shows 3GPP \ac{TDL}-A channel image detection example at 20dB \ac{SNR} where \ac{Uni-DNN} arch. C achieves a \ac{BER} of $7 \times 10^{-4}$ while \ac{LS} and perfect-\ac{CSI} \ac{MMSE} achieve $7 \times 10^{-3}$ and $3 \times 10^{-3}$ respectively. Also, we can see that feeding the classification as flattened 2D-grid input in arch. B and employing \ac{CNN} in arch. C have further enhanced the \ac{BER} performance compared to arch. A. This improvement can be linked to \ac{Uni-DNN} arch. B and C's per-channel weights assignment discussed earlier in section \ref{Uni-DNN section}. Although the channel classifier facilitated performance enhancement for all \ac{Uni-DNN} architectures at high \ac{SNR} in the case of 3GPP \ac{TDL}-A channel, such enhancement is less apparent at low \ac{SNR} due to misclassification of the channel as WINNER II since the two channels are correlated.

\begin{figure}[ht]
\centering
\includegraphics[width=0.79\columnwidth]{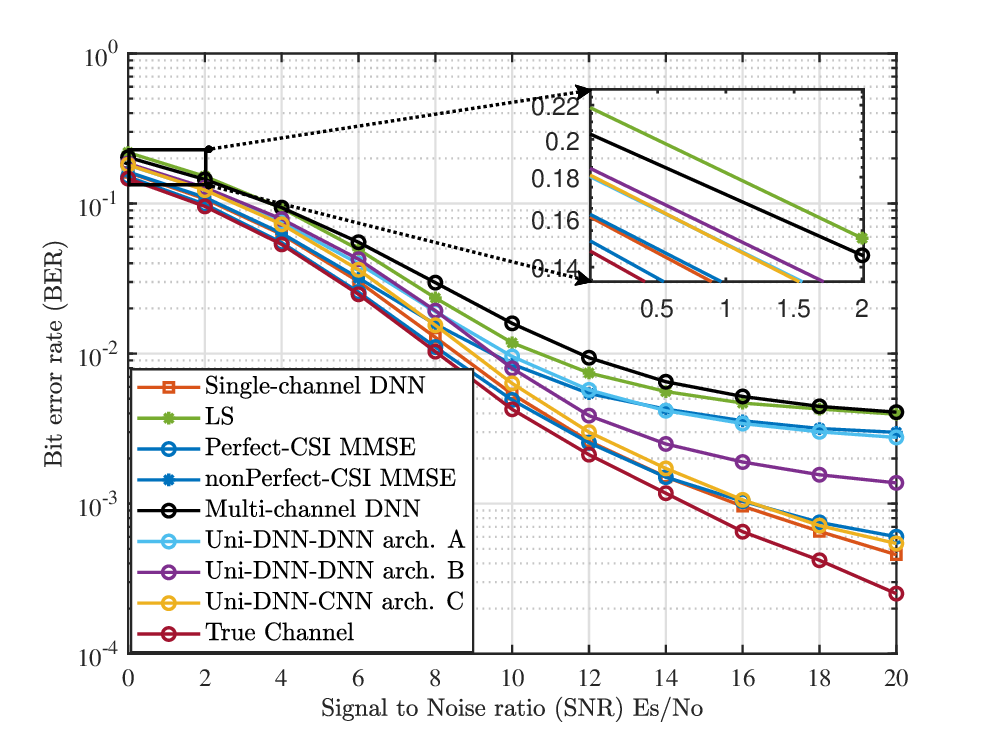}
\caption[32-pilots 3GPP TDL-A channel BER performance vs. SNR for conventional and DL methods comparison.]{32-pilots 3GPP TDL-A channel BER performance vs. SNR for conventional and DL methods comparison.}
\label{fig_1.2b}
\end{figure}%

\begin{figure}[ht]
\centering
\includegraphics[width=0.79\columnwidth]{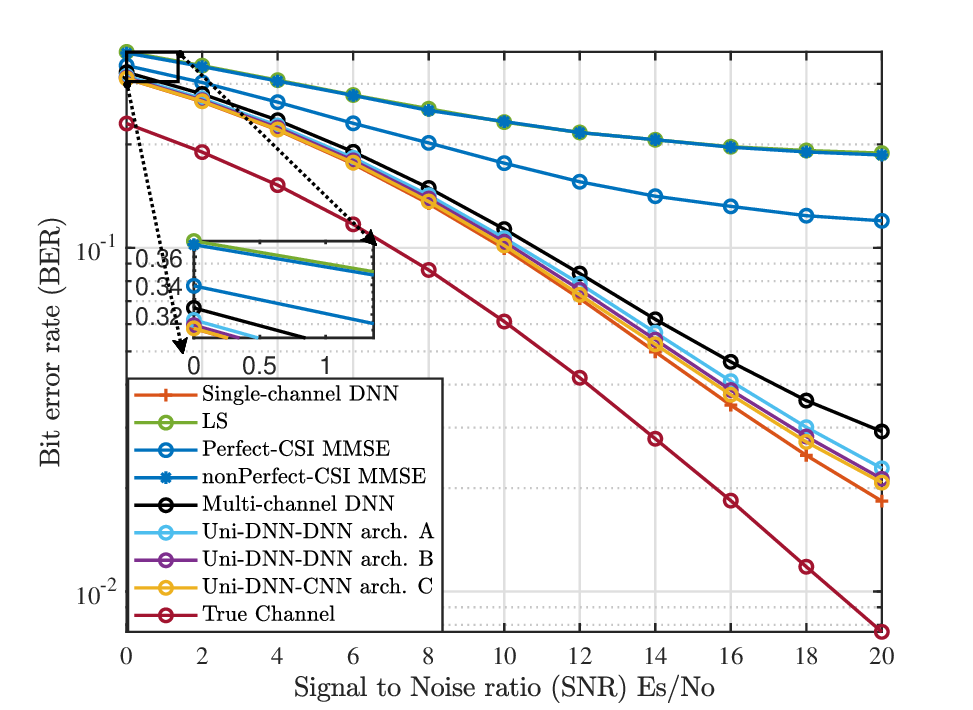}
\caption[8-pilots Rician channel BER performance vs. SNR for conventional and DL methods comparison.]{8-pilots Rician channel BER performance vs. SNR for conventional and DL methods comparison.}
\label{fig_1.4}
\end{figure}
\begin{figure}[ht]
\centering
\includegraphics[width=\columnwidth]{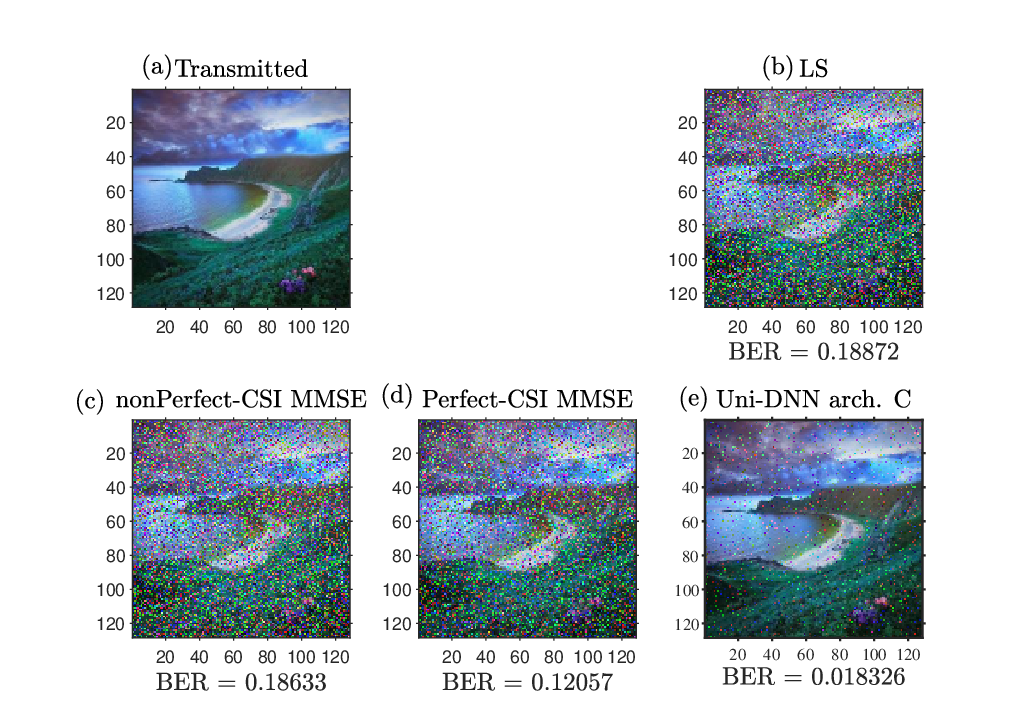}
\caption[Rician channel (a) Transmitted, (b) LS, (c) non-perfect, (d) perfect MMSE and (e) DNN equalised image at high SNR scenario namely 20 dB.]{Rician channel (a) Transmitted, (b) LS, (c) non-perfect, (d) perfect MMSE and (e) DNN equalised image at 20 dB.}
\label{fig_1.5}
\end{figure}
\begin{figure}[ht]
\centering
\includegraphics[width=0.79\columnwidth]{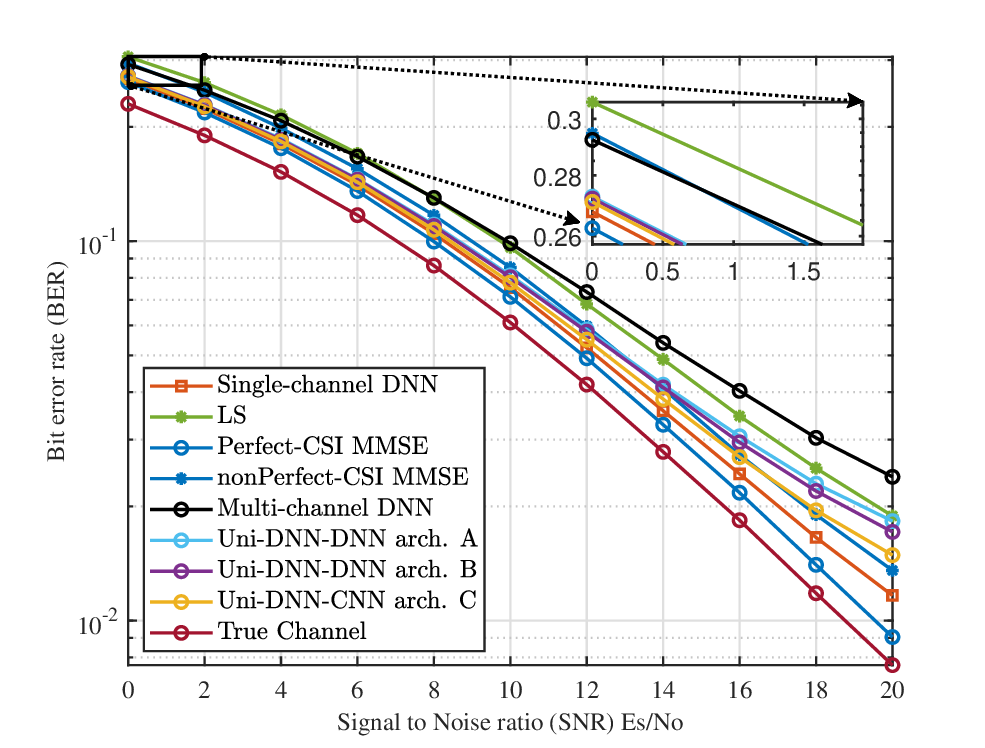}
\caption[32-pilots Rician channel BER performance vs. SNR for conventional and DL methods comparison.]{32-pilots Rician channel BER performance vs. SNR for conventional and DL methods comparison.}
\label{fig_1.4b}
\end{figure}

For the simulated Rician channel model, we can see from Fig.~\ref{fig_1.4} and Fig.~\ref{fig_1.5} that due to an insufficient number of pilots to estimate and interpolate a highly selective channel with 6 taps the performance of \ac{LS}, perfect and non-perfect-\ac{CSI} \ac{MMSE} are poor and much inferior compared to \ac{DL} models. Similar observations to the previous channel model can be seen but unlike the 3GPP \ac{TDL}-A channel, the Rician channel has been correctly classified in high and low \ac{SNR} which maintained the \ac{Uni-DNN} architectures' \ac{BER} advantage across the whole \ac{SNR} range. This accuracy can be attributed to the presence of a \ac{LOS} component in the Rician channel, which exhibits a non-random pattern that is distinguishable compared to other channels. Overall, increasing the number of pilots reduces the performance gap between conventional methods and \ac{DL} models, as well as closing the gap between all \ac{DL} models and true channel \ac{BER} performance curve as in Fig.~\ref{fig_1.2b} and Fig.~\ref{fig_1.4b}.


\subsection{Experimental Computational Complexity Comparison}
\label{Experimental complexity comparison}

In addition to the analytical computational complexity analysis discussed in subsection \ref{Analytical complexity comparison}, in this section, the inference time relative to the \ac{LS} channel estimator for each method is shown in Table \ref{Practical Complexity analysis table runtime}. The experimental relative runtime results are obtained on hardware running Python on NVIDIA RTX-3060 GPU and $12^{th}$ generation i7 Intel CPU with 16GB of RAM. we can see that the experimental results are consistent with the analytical results in Table \ref{infernce_complexity}.

\begin{table}[ht]
\caption{Model inference relative run-time analysis.}
\label{Practical Complexity analysis table runtime}
\begin{tabular}[t]{>{\centering\arraybackslash}p{\mylength} >{\centering\arraybackslash}p{\mylength}}
\hline
Model/Parameter & Inference Relative runtime\\
\hline
\ac{LS}$^{*}$ & $T_{LS} = 5.16 \times 10^{-7}\ sec$ \\
\ac{MMSE} & $~6,243\ T_{LS}$ \\
Single-channel & $~63\ T_{LS}$ \\
Multi-channel & $~63\ T_{LS}$ \\
Uni-Arc-A$^{**}$ &  $~88\ T_{LS}$ \\
Uni-Arc-B$^{**}$ & $~92\ T_{LS}$ \\
Uni-Arc-C$^{**}$ & $~151\ T_{LS}$ \\
\hline
\end{tabular}
\\ \raggedright\footnotesize{ \ \ \ \ \ $^*$Excluding linear interpolation \ \ **Including channel classifier}\\
\end{table}%

\section{Conclusion}

In this article, three universal cascaded \ac{DL}-based models are proposed for joint channel estimation and signal detection. These models have shown better adaption to various channel models compared to \ac{DL} models employing one \ac{DNN} trained on multi-channels. Also, \ac{Uni-DNN} architecture C has managed to minimize the gap in \ac{BER} performance with single-channel trained \ac{DL} models outperforming conventional estimation methods for various channel models without the requirement of re-training. \ac{Uni-DNN} models have faster inference capability compared to \ac{MMSE} estimators making it attractive for practical real-time applications as a universal basestation receiver. Simulating the proposed \ac{Uni-DNN} on more complex scenarios such as pilotless or multiple-input-multiple-output \ac{OFDM} or implementing it on hardware to verify the simulation results can be done as future work.

\bibliographystyle{IEEEtran}
\bibliography{references}

\end{document}